\documentclass{article}
\usepackage[utf8]{inputenc}
\usepackage{amsmath}
\usepackage{amsfonts}
\usepackage{amssymb}
\usepackage{mathtools}
\usepackage{stmaryrd}
\usepackage{amsthm}
\usepackage{tikz-cd}

\makeatletter
\DeclareFontEncoding{LS1}{}{}
\makeatother
\DeclareFontSubstitution{LS1}{stix}{m}{n}
\DeclareSymbolFont{operatorsstix}{LS1}{stix}{m}{n}
\DeclareSymbolFont{arrows1stix}{LS1}{stixsf}{m}{n}
\DeclareSymbolFont{arrows2stix}{LS1}{stixsf}{m}{it}
\DeclareMathSymbol{\upand}{\mathbin}{operatorsstix}{"C4}
\DeclareMathSymbol{\nvrightarrow}             {\mathrel}{arrows1stix}{"F6}
\DeclareMathSymbol{\lefttail}                 {\mathrel}{arrows2stix}{"B2}

\theoremstyle{definition}
\newtheorem{defi}{Definition}[section]
\newtheorem{rema}[defi]{Remark}
\newtheorem{nota}[defi]{Notation}
\newtheorem{theo}[defi]{Theorem}
\newtheorem{prop}[defi]{Proposition}
\newtheorem{lemm}[defi]{Lemma}
\newtheorem{conj}[defi]{Conjecture}

\begin{document}

\begin{center}
    \huge\textbf{Concurrent Strategies on Games with Algebras}
\end{center}
\vspace{1cm}
\begin{center}
    \large{Sacha Huriot-Tattegrain,\\
M1 of Computer Science Jacques Herbrand, ENS Paris-Saclay}
\end{center}
\begin{center}
    \large{under the supervision of Glynn Winskel,\\ Department of Computer Science and Technology,\\ University of Cambridge, Cambridge, United-Kingdom}
\end{center}
\begin{center}
    \large\today
\end{center}
\vspace{1cm}
\tableofcontents
\newpage
\section{Introduction}
\subsection{Context of the subject}
Concurrency theory is the sub-field of computer science which uses structures to represent the ability of different parts of an algorithm or a system to be used independently - usually in any order - to solve a specific problem. Applied to computer process, concurrency helps to plan parallel execution, which can significantly improve overall speed of execution. Applied to Algorithmics, concurrency refers to the degree to which a certain program can be decomposed into components which can be executed is (almost) any order. One of the first concurrency tools was proposed by Carl Adam Petri's seminal work on Petri nets in the early 1960s. In the years since, a wide variety of formalism have been developed for modeling and reasoning about concurrency.\par
Event structures are a model used in concurrency \cite{ecsym}, where processes are events constrained by relations of consistency and causality. Their study first arose on some foundational questions in denotational semantics, and since then they have been developed as a model in their own right and for certain applications they are easier and less clumsy to use than Petri nets to which they are closely related. When looking at games played on event structures - to usually simulate one process against the rest of the system - the set of strategies for the player forms interesting structures like bi-categories \cite{Cstrat}. This game theoretic approach to the study of event structures allows to easily add extra operations and structure to the strategies, with quantum operators \cite{quantum} or imperfect information \cite{imperfectinfo} for example. It also provides different ways to study the optimality of the concurrent algorithms, e.g. by introducing payoff functions \cite{payoff}.\par
Concurrent games with algebra have so far not been investigated thoroughly, where
Player and Opponent can make use of algebraic properties to employ more complex probabilistic concurrent strategies. In a situation where Player and Opponent make concurrent moves according to a game with algebra, they can use a richer logic with terms and variables linked to the events. The first part of this report involves introducing the formalism and adding algebraic structure, then presenting a new concurrent game that can encompass the algebra to be able to use known properties of event structures. The second part consists in trying to explore and study composition of strategies on these games and the classic extensions of concurrent games like neutral events or imperfect information.\par
The initial purpose of the internship was first to extend the classic results (like the bicategory structure) of concurrent games to games with algebra and in a second phase to include imperfect information and be able to study optimality of the strategies with a payoff function. However we discovered that strategies with algebra don't compose as well as classic ones and the situation of the internship made significant progress harder to achieve. This report thus also tries to present the different paths of inquiry we took from working on concurrent games with algebras.
\newpage
\subsection{Preliminary definitions}
Event structures are a model of computational processes. They represent a system, as a set of event occurrences with relations to express how events causally depend on others, or exclude other events from occurring. In one of their simpler forms they consist of a set of events on which there is a consistency relation expressing when events can occur together in a history and a partial order of causal dependency.
\begin{defi}Event Structure\cite{ecsym}\\
Let $E$ be a set of $events$ which are partially ordered by the causal dependency relation $\leq$, and let $Con$ be a nonempty consistency relation consisting of finite subsets of $E$ such that:
\begin{align*}
    \forall e\in E,\ \{e'\mid e'\leq e\}&\text{ is finite and }\{e\}\in Con\\
    Y\subseteq X\wedge X\in &\ Con\Rightarrow Y\in Con\\
    X\in Con\wedge e\leq e'\wedge e&'\in X\Rightarrow X\cup\{e\}\in Con
\end{align*}
Then $(E,\leq,Con)$ is an event structure. We say that events $e$ and $e'$ are $concurrent$, and write $e\ co\ e'$, if $\{e, e'\}\in Con\wedge e\not\leq e'\wedge e'\not\leq e$. Concurrent events can occur together, independently of each other. The relation of immediate dependency $e\rightarrow e'$ means $e$ and $e'$ are distinct and $e\leq e'$ with no events in between.
\end{defi}
An event structure represents a process. A configuration is the set of all events which may have occurred by some stage, or history, in the evolution of the process. According to our understanding of the consistency relation and causal dependency relations a configuration should be consistent and such that if an event appears in a configuration then so do all the events on which it causally depends.
\begin{defi}Configurations\\
A configuration of an event structure $E$ is a subset of events $x\subseteq E$ such that:
\begin{align*}
    &\forall X\subseteq x,\ X\text{ finite}\Rightarrow X\in Con\\
    \forall e&,e'\in E,\ e'\leq e\wedge e\in x\Rightarrow e'\in x
\end{align*}
The set of all configurations is denoted by $\mathcal{C}^\infty(E)$ and the finite ones by $\mathcal{C}(E)$. We later use $x\stackrel{e}{\lefttail}$ , expressing that event $e$ is enabled at configuration $x$, i.e. $x\cup\{e\}$ is a also configuration of $E$.
\end{defi}
Finally let's look at the $maps$ of event structures that will serve to define strategies later.
\begin{defi}Maps of event structures\\
Let $E$ and $E'$ be event structures. A (partial) map of event structures is a (partial) function on events $f: E\rightarrow E'$ such that for all configuration $x\in\mathcal{C}^\infty(E)$ its direct image is a configuration $fx\in\mathcal{C}^\infty(E)$ and
\begin{center}
    $\forall e_1,e_2\in x,\ f(e_1)=f(e_2)\Rightarrow e_1=e_2$
\end{center}
\end{defi}
The map expresses how the occurrence of an event $e\in E$ induces the coincident occurrence of the event $f(e)\in E'$ whenever it is defined. The map $f$ respects the instantaneous nature of events: two distinct event occurrences which are consistent with each other cannot both coincide with the occurrence of a common event in the image.
\section{Games and Strategies}
\subsection{Adjoining an algebra to a game}
We shall represent a concurrent game as an event structure with polarity, comprising an event structure together with a polarity function $pol:E\rightarrow\{+,-\}$ assigning a polarity + or — to its events $E$. The events correspond to (occurrences of) moves. The two polarities +/— express dichotomy: Player/Opponent, Process/Environment, or Ally/Enemy. Maps of event structures with polarity are maps of event structures which preserve polarity.
\begin{defi}Strategy on an event structure with polarity\\
Let $f:S\rightarrow E$ be a total map of event structures, then we call $f$ a strategy for the concurrent game $(E,pol)$ if and only if:
\begin{itemize}
    \item $f$ is \textbf{receptive}
    \begin{align*}
        \forall x\in\mathcal{C}^\infty(S),\ e\in E,\ fx\stackrel{e}{\lefttail}\wedge pol(e)=—\Rightarrow\exists!s\in S\text{ s.t. }x\stackrel{s}{\lefttail}\wedge f(s)=e
    \end{align*}
    \item $f$ is \textbf{innocent}
    \begin{align*}
        &s\rightarrow s'\ \wedge\ pol(s')=-\Rightarrow\sigma(s)\rightarrow\sigma(s')\\
        &s\rightarrow s'\ \wedge\ pol(s)=+\Rightarrow\sigma(s)\rightarrow\sigma(s')
\end{align*}
\end{itemize}
\end{defi}
\begin{defi}Basic Operations on games
\begin{itemize}
    \item Dual polarity\\
    The dual, $E^\bot$, of an event structure with polarity $E$ comprises a copy of the event structure $E$ but with a reversal of polarities. Write $\overline{e}\in E^\bot$ for the event complementary to $e\in E$ and vice versa.
    \item Simple parallel composition\\
    Let $(A, \leq_A ,Con_A,pol_A)$ and $(B,\leq_B,Con_B,pol_B)$ be event structures with polarity. Their parallel composition, written $A||B$, has events in $(\{1\}\times A)\cup(\{2\}\times B)$ with their polarities unchanged. Moreover, the only relations of causal dependency are given by $(1,a)\leq (1,a')$ iff $a\leq_A a'$ and $(2,b)\leq(2,b')$ iff $b\leq_B b'$; a subset of events $C$ is consistent in $A||B$ iff $\{a\mid(1,a)\in C\}\in Con_A$ and $\{b\mid (2,b)\in C\}\in Con_B$.
\end{itemize}
\end{defi}
It is shown in \cite{bicategory} that - with these strategies as maps and event structures as objects - general concurrent games form a bicategory. We hence looked to find the same kind of structure after adding algebras over event structures. The kind of algebraic structure we're using is a \textit{relational many-sorted algebra} which is useful in simulating a basic and general logic for strongly-typed programming languages. 
\begin{defi}A relational many-sorted algebra is given by
\begin{align*}
    \mathcal{A}=(|\mathcal{A}|,sort,\Sigma,(–)\mapsto(–)^\mathcal{A})
\end{align*}
where $|\mathcal{A}|$ is the set of elements and the sorts are given by the sort function $sort:|A|\rightarrow Sorts$. The relational operators $R$ are in the signature $\Sigma=\bigcup_w\Sigma_w$ with the interpretation function $R\mapsto R^\mathcal{A}$ such that $R\in\Sigma_{s_1s_2...s_k}\Rightarrow R^\mathcal{A}\subseteq sort^{-1}(s_1)\times sort^{-1}(s_2)\times...\times sort^{-1}(s_k)$.
\end{defi}
\begin{defi} Game on a relational many-sorted algebra\\
W.r.t. a relational many-sorted algebra $\mathcal{A}$, with signature $\Sigma$, define a $\mathcal{A}$-game to comprise
\begin{align*}
    (\mathcal{A}, E, pol, var)
\end{align*}
where $var : E\rightarrow Var$ where $Var$ a countable set of sorted variables for $\mathcal{A}$ and $(E, pol)$ is an event structure with polarity satisfying
\begin{align*}
    \forall x\in\mathcal{C}(E),\ x\stackrel{e}{\lefttail}\wedge x\stackrel{e'}{\lefttail}\wedge &pol(e),pol(e')=(+,-)\Rightarrow x\cup\{e,e'\}\in\mathcal{C}(E)\tag{\textbf{race-free}}
\end{align*}
and
\begin{align*}
    \forall e, e'\in E,\ var(e)=var(e')\Rightarrow \neg(e\ co\ e')\tag{\textbf{no-overlap}}
\end{align*}
\end{defi}\par
A \textit{race-free} event structure only allows immediate conflict between events of the same polarity; there can be no races between Player and Opponent moves. It is a standard property for adversarial games.\par
As a consequence of the axiom (\textbf{no-overlap}), we obtain that the set of events in a configuration $x$ of $E$ having a common variable is totally ordered w.r.t. causal dependency $\le$.\par
For $\alpha\in Var,\ x\in\mathcal{C}^\infty(E)$ write $latest(\alpha,x)$ for the $\le$-greatest $e\in x$ such that $var(e)=\alpha$, if such exists. It may not exists through $\alpha\not\in var(x)$ or through $x$ being infinite and including an infinite chain of events labelled by $\alpha$.
\begin{rema}
The event structure of an $\mathcal{A}$-game depends only on the signature and variables and not on $\mathcal{A}$ itself. The precise algebra $\mathcal{A}$ plays a role in the definition of $\mathcal{A}$-strategy and in the $\mathcal{A}$-expansion of a game used in the definition of composition of strategies and copycat strategy.
\end{rema}
A winning condition for the game is an assertion in the free logic over signature $\Sigma$ with the variables in $Var$ and the names for elements in $|\mathcal{A}|$ adjoined (by using a free logic we avoid indexing winning conditions by configurations of the game).
\begin{defi}Free logic w.r.t. $\mathcal{A}$\\
Its terms are either variables or elements (which carry sorts):
\begin{align*}
    t::=\alpha\in Var\ \mid\ a\in|\mathcal{A}|
\end{align*}
Its assertions are:
\begin{align*}
    \phi::&=\ R(t_1,...,t_k)\ \mid\ t_1=t_2\ \mid\ \mathbb{E}(t)\\
    &\ \mid\ \phi_1\wedge \phi_2\ \mid\ \phi_1\vee \phi_2\ \mid\ \neg\phi\ \mid\ \forall\alpha.\phi\ \mid\ \exists\alpha.\phi\\
    &\ \mid\ \bigwedge_{i\in I}\phi_i\ \mid\ \bigvee_{i\in I}\phi_i
\end{align*}
with $t_i$ and $t$ ranging over terms, adequate sorts and arity for $R$ (i.e. $R\in\Sigma_w$ and $sort(t_i)=s_i$ where $w=s_1...s_k$) and $I$ countable.
\end{defi}\par
\begin{nota}For a term or an assertion in the free logic w.r.t. $\mathcal{A}$, define its size as the following ordinal number:
\begin{align*}
    size(a)=size(\alpha)=1\quad size(\mathbb{E}(t))&=2\quad size(t_1=t_2)=3\\
    size(R(t_1,...,t_k))&=k+1\\
    size(\neg\phi)=size(\forall\alpha.\phi)=size(\exists\alpha.\phi)&=size(\phi)+1\\
    size(\phi_1\wedge\phi_2)=size(\phi_1\vee\phi_2)&=sup\{size(\phi_1),size(\phi_2)\}+1\\
    size(\bigwedge_{i\in I}\phi_i)=size(\bigvee_{i\in I}\phi_i)&=sup\{size(\phi_i)\mid i\in I\}+1
\end{align*}
Remark that for any assertion $\phi$ we have $size(\phi)=size(\phi[a/\alpha])$.
\end{nota}
We give the semantics shortly w.r.t. an $\mathcal{A}$-strategy.
\begin{defi}Strategy in an $\mathcal{A}$-game\\
An $\mathcal{A}$-strategy in an $\mathcal{A}$-game $E$ comprises a total map of event structures with polarity
\begin{align*}
    &S\xrightarrow{\text{  $\sigma$ }}E\text{ with}\\
    inst:S\rightarrow|\mathcal{A}|\text{ s.t. }&\forall s\in S,\ sort(inst(s))=sort(var(\sigma(s))) 
\end{align*}
which is \textbf{receptive}:
\begin{align*}
    &\forall x\in\mathcal{C}^\infty(S),\ \forall e\in E,\ \sigma x\stackrel{e}{\lefttail}\ \&\ pol(e)=-\ \&\ var(e)=\alpha,\\
    &\forall a\in|\mathcal{A}|,\ sort(a)=sort(\alpha),\\
    &\exists!s\in S,\ x\stackrel{s}{\lefttail}\ \&\ \sigma(s)=e\ \&\ inst(s)=a
\end{align*}
and \textbf{innocent/courteous}:
\begin{align*}
    &s\rightarrow s'\ \&\ pol(s')=-\Rightarrow\sigma(s)\rightarrow\sigma(s')\\
    &s\rightarrow s'\ \&\ pol(s)=+\Rightarrow\sigma(s)\rightarrow\sigma(s')
\end{align*}
\end{defi}\par
Such an $\mathcal{A}$-strategy is winning w.r.t. a winning condition $W$ iff\\ $\forall x\in\mathcal{C}^\infty(S),\ x\text{ is $+$-maximal}\Rightarrow x\models W$. This definition requires the semantics of the free logic for $x\models W$. A term or assertion will be associated with the set of configurations of S at which is meaningful, and in the case of assertions true.
\begin{nota} For $x\in\mathcal{C}^\infty(S)$ write
\begin{align*}
    last(x)&:=inst\{s\in x\mid\exists\alpha\in Var,\ \sigma(s)=latest(\alpha,\sigma x)\}\\
    &\subseteq |\mathcal{A}|
\end{align*}
\end{nota}
\begin{defi}Semantics of free logic\\
Let $\sigma,\ inst$ be an $\mathcal{A}$-strategy, define the semantics of terms as:
\begin{align*}
    &\llbracket\alpha\rrbracket=\{x\in\mathcal{C}^\infty(S)\mid\alpha\in var(\sigma x)\}\\
    &\llbracket a\rrbracket=\{x\in\mathcal{C}^\infty(S)\mid a\in last(x)\}
\end{align*}\par
Then for a term $t$ and $x\in\llbracket t\rrbracket$ define $t[x]$ as follows:\\
If $t$ is $\alpha\in Var$, then $t[x]=inst(s)$ where $s$ is the unique element of $x$ for which $\sigma(s)=latest(\alpha,\sigma x)$. If $t$ is $a\in|\mathcal{A}|$, then $t[x]=a$.\par
Define the semantics of assertions by induction on length as:
\begin{align*}
    \llbracket R(t_1,...,t_k)\rrbracket&=\{x\in\mathcal{C}^\infty(S)\mid x\in\llbracket t_1\rrbracket\cap...\cap\llbracket t_k\rrbracket\ \&\ (t_1[x],...,t_k[x])\in R^\mathcal{A}\}\\
    \llbracket t_1=t_2\rrbracket&=\{x\in\mathcal{C}^\infty(S)\mid x\in\llbracket t_1\rrbracket\cap\llbracket t_k\rrbracket\ \&\ t_1[x]=_\mathcal{A}t_2[x]\}\\
    \llbracket\mathbb{E}(t)\rrbracket&=\llbracket t\rrbracket\\
    \llbracket\phi_1\wedge\phi_2\rrbracket&=\llbracket\phi_1\rrbracket\cap\llbracket\phi_2\rrbracket\\
    \llbracket\phi_1\vee\phi_2\rrbracket&=\llbracket\phi_1\rrbracket\cup\llbracket\phi_2\rrbracket\\
    \llbracket\neg\phi\rrbracket&=\mathcal{C}^\infty(S)\setminus\llbracket\phi\rrbracket\\
    \llbracket\forall\alpha.\phi\rrbracket&=\{x\in\mathcal{C}^\infty(S)\mid\forall a\in|\mathcal{A}|,\ sort(a)=sort(\alpha)\ \&\ a\in last(x)\Rightarrow x\in\llbracket\phi[a/\alpha]\rrbracket\}\\
    \llbracket\exists\alpha.\phi\rrbracket&=\{x\in\mathcal{C}^\infty(S)\mid\exists a\in|\mathcal{A}|,\ sort(a)=sort(\alpha)\ \&\ a\in last(x)\ \&\ x\in\llbracket\phi[a/\alpha]\rrbracket\}\\
    \llbracket\bigwedge_{i\in I}\phi_i\rrbracket&=\bigcap_{i\in I}\llbracket\phi_i\rrbracket\\
    \llbracket\bigvee_{i\in I}\phi_i\rrbracket&=\bigcup_{i\in I}\llbracket\phi_i\rrbracket
\end{align*}
Finally write $x\models\phi$ iff $x\in\llbracket\phi\rrbracket$.
\end{defi}
\subsection{Relation between games with and without algebra}
We first expend the game operations before introducing a way to construct a big concurrent game without algebra from a game with algebra. Once some preliminary results are provided we present our main theorem, that $\mathcal{A}$-strategies can be simulated by classic strategy on this bigger concurrent game.
\begin{defi}Dual\\
Define the dual of an $\mathcal{A}$-game $(E,\mathcal{A},var,W)$ as:
\begin{align*}
    (E,\mathcal{A},var,W)^\bot=(E^\bot,\mathcal{A},var,\neg W)
\end{align*}
\end{defi}
\begin{defi}Parallel composition\\
Consider the $\mathcal{A}$ and $\mathcal{B}$-games:
\begin{align*}
    (E,\mathcal{A},var_E:E\rightarrow Var_\mathcal{A},W_E)\text{ and }(F,\mathcal{B},var_F:F\rightarrow Var_\mathcal{B},W_F)
\end{align*}
Define their parallel composition as the following $\mathcal{A}||\mathcal{B}$-game:
\begin{align*}
    (E,\mathcal{A},var_E,W_E)\ \upand\ (F,\mathcal{B},var_F,W_F)=(E||F,\mathcal{A}||\mathcal{B},var_E||var_F,W_E\vee W_F)
\end{align*}
The many-sorted algebra $\mathcal{A}||\mathcal{B}$ has for set of elements, for sorts, for signature and for set of relations the respective disjoint unions of the ones of $\mathcal{A}$ and $\mathcal{B}$.\\
The map $var_E||var_F:E||F\rightarrow Var_{\mathcal{A}||\mathcal{B}}$ is defined s.t. $(var_E||var_F)_E=var_E$ and $(var_E||var_F)_F=var_F$.
\end{defi}
Given an $\mathcal{A}||\mathcal{B}$-strategy $\sigma:S\rightarrow E||F$ (and $inst_\sigma$), we can form the two defined parts $\sigma_E$ and $\sigma_F$:
\begin{center}
\begin{tikzcd}
S_E \arrow[d, "\sigma_E"', dashed] & S \arrow[d, "\sigma"'] \arrow[l, "p_E"', harpoon', dashed] \arrow[r, "p_F", harpoon, dashed] & S_F \arrow[d, "\sigma_F", dashed] \\
E                         & E||F \arrow[r, harpoon] \arrow[l, harpoon']                                            & F                       
\end{tikzcd}
\end{center}
It's easy to check that $\sigma_E$ and $\sigma_F$ are strategies (they needn't be determinate even though $\sigma$ is), and moreover $\mathcal{A}$ and $\mathcal{B}$-strategies respectively by inheriting instantiations from $inst_\sigma$ (we have $inst_{S_E}=inst_{\sigma_E}$ and $inst_{S_F}=inst_{\sigma_F}$). For $x\in\mathcal{C}^\infty(S)$ write $x_E:=p_E x\in\mathcal{C}^\infty(S_E)$ and $x_F:=p_F x\in\mathcal{C}^\infty(S_F)$.

\begin{lemm}\label{lma2}
$x\models_\sigma W_E\vee W_F$  iff  ($x_E\models_{\sigma_E}W_E$ or $x_F\models_{\sigma_F}W_F$)
\end{lemm}
$Proof.\ $We have:
 \begin{align*}
    x\models_\sigma W_E\vee W_F&\iff x\in\llbracket W_E\vee W_F\rrbracket_\sigma\iff x\in\llbracket W_E\rrbracket_\sigma\cup\llbracket W_F\rrbracket_\sigma\\
    &\iff x\in\llbracket W_E\rrbracket_\sigma\text{ or }x\in\llbracket W_F\rrbracket_\sigma\iff x\models_\sigma W_E\text{ or }x\models_\sigma W_F
\end{align*}
Now let's show $x\models_\sigma W_E\iff x_E\models_{\sigma_E}W_E$ by induction on the size of $W_E$.
\begin{itemize}
    \item If $W_E=\mathbb{E}(\alpha)$ s.t. $\alpha\in Var_\mathcal{A}$ then $\llbracket W_E\rrbracket_\sigma=\llbracket\alpha\rrbracket_\sigma$. Remark that we have $\alpha\not\in Var_\mathcal{B}=Dom(var_F)$ and $(\sigma x)_E=\sigma_E x_E$ so:
    \begin{align*}
        x\in\llbracket\alpha\rrbracket_\sigma&\iff\alpha\in(var_E||var_F)(\sigma x)\\
        &\iff\alpha\in var_E((\sigma x)_E)\\
        &\iff\alpha\in var_E(\sigma_E x_E)\\
        &\iff x_E\in\llbracket\alpha\rrbracket_{\sigma_E}
    \end{align*}
    \item If $W_E=\mathbb{E}(a)$ s.t. $a\in|\mathcal{A}|$ then $\llbracket W_E\rrbracket_\sigma=\llbracket a\rrbracket_\sigma$. Remark that we have $inst_\sigma(s)\in|\mathcal{A}|\iff s\in S_E\iff \sigma(s)=(1,\sigma_E(s))\iff(var_E||var_F)(\sigma(s))=var_E(\sigma_E(s))\in Var_\mathcal{A}$ by property of $inst$. What's more, $\alpha\in Var_\mathcal{A}\Rightarrow latest(\alpha,\sigma x)=latest(\alpha,(\sigma x)_E)$ so:
    \begin{align*}
        x\in\llbracket a\rrbracket_\sigma&\iff a\in last_\sigma(x)\\
        &\iff a=inst_\sigma(s)\text{ s.t. } \exists\alpha\in Var_{\mathcal{A}||\mathcal{B}},\ \sigma(s)=latest(\alpha,\sigma x)\\
        &\iff a=inst_{\sigma_E}(s)\text{ s.t. } \exists\alpha\in Var_\mathcal{A},\ \sigma(s)=latest(\alpha,\sigma x)\\
        &\iff a=inst_{\sigma_E}(s)\text{ s.t. } \exists\alpha\in Var_\mathcal{A},\ \sigma_E(s)=latest(\alpha,(\sigma x)_E)\\
        &\iff a=inst_{\sigma_E}(s)\text{ s.t. } \exists\alpha\in Var_\mathcal{A},\ \sigma_E(s)=latest(\alpha,\sigma_E x_E)\\
        &\iff a\in last_{\sigma_E}(x_E)\\
        &\iff x_E\in\llbracket a\rrbracket_{\sigma_E}
    \end{align*}
\end{itemize}
Remark that with these two results the definitions of $t[x]$ and $t[x_E]$ coincide for any term $t$ of the free logic w.r.t. $\mathcal{A}||\mathcal{B}$ as long as $t$ is also one w.r.t. $\mathcal{A}$. It follows by induction that $x\in\llbracket R(t_1,...,t_k)\rrbracket_\sigma\iff x_E\in\llbracket R(t_1,...,t_k)\rrbracket_{\sigma_E}$ and $x\in\llbracket t_1=t_2\rrbracket_\sigma\iff x_E\in\llbracket t_1=t_2\rrbracket_{\sigma_E}$ because $R^\mathcal{A}$ and $=_\mathcal{A}$ are independent from the strategy.\\ The induction cases for $\wedge$, $\vee$, $\neg$, $\bigwedge_I$ and $\bigvee_I$ are trivial by definition of the size of an assertion. This leaves the cases of quantifiers:
\begin{itemize}
    \item If $W_E=\forall\alpha.\phi$ with $\alpha\in Var_\mathcal{A}$ and by induction hypothesis $size(\phi')<size(W_E)\Rightarrow (x\models_\sigma\phi'\iff x_E\models_{\sigma_E}\phi')$. Remark that $\alpha\in Var_\mathcal{A}\ \&\ sort(a)=sort(\alpha)\Rightarrow a\not\in|\mathcal{B}|$, then from the case of $\mathbb{E}(a)$ we have $a\in|\mathcal{A}|\Rightarrow (a\in last_\sigma(x)\iff a\in last_{\sigma_E}(x_E))$ and finally $size(\phi[a/\alpha])=size(\phi)\le size(\forall\alpha.\phi)$ hence $x\models_\sigma\phi[a/\alpha]\iff x_E\models_{\sigma_E}\phi[a/\alpha]$, so:
    \begin{align*}
        x\in\llbracket\forall\alpha.\phi\rrbracket_\sigma&\iff(\forall a\in|\mathcal{A||B}|,\ sort(a)=sort(\alpha)\ \&\ a\in last_\sigma(x)\Rightarrow x\in\llbracket\phi[a/\alpha]\rrbracket_\sigma)\\
        &\iff(\forall a\in|\mathcal{A}|,\ sort(a)=sort(\alpha)\ \&\ a\in last_\sigma(x)\Rightarrow x\in\llbracket\phi[a/\alpha]\rrbracket_\sigma)\\
        &\iff(\forall a\in|\mathcal{A}|,\ sort(a)=sort(\alpha)\ \&\ a\in last_{\sigma_E}(x_E)\Rightarrow x\in\llbracket\phi[a/\alpha]\rrbracket_\sigma)\\
        &\iff(\forall a\in|\mathcal{A}|,\ sort(a)=sort(\alpha)\ \&\ a\in last_{\sigma_E}(x_E)\Rightarrow x_E\in\llbracket\phi[a/\alpha]\rrbracket_{\sigma_E})\\
        &\iff x_E\in\llbracket\forall\alpha.\phi\rrbracket_{\sigma_E}
    \end{align*}
    \item Same case for $\exists\alpha.\phi$.
\end{itemize}
By the same induction, $x\models_\sigma W_F\iff x_F\models_{\sigma_F}W_F$.
\begin{flushright}\qed\end{flushright}
\begin{defi}Strategy from $E$ to $F$\\
A strategy form $E$ to $F$ is an $\mathcal{A}||\mathcal{B}$-strategy $\sigma$ in the game
\begin{align*}
    E^\bot\upand F=(E^\bot||F,\mathcal{A}||\mathcal{B},var_E||var_F,W_E\Rightarrow W_F)
\end{align*}
which is winning. Write $\sigma:E\nvrightarrow F$.
\end{defi}
Now we define a way to obtain an event structure with polarity from an $\mathcal{A}$-game.
\begin{defi}$\mathcal{A}$-expansion\\
Define $expn(E,\mathcal{A},var)$ to have events $(e,\epsilon)$ where\\ $\epsilon:[e]\rightarrow|\mathcal{A}|$ s.t. $sort(\epsilon(e'))=sort(var(e'))$ for all $e'\in[e]$.\\
Write $(e',\epsilon')\le(e,\epsilon)$ iff $e'\le e\ \&\ \epsilon'=\epsilon\upharpoonright[e']$.\\
Write $X\in Con_{expn(E)}$ iff $\Pi_1X\in Con_E$ and $\forall(e,\epsilon),(e',\epsilon')\in X,\ e=e'\Rightarrow\epsilon=\epsilon'$.\\
The event structure $expn(E,\mathcal{A},var)$ will sometimes be referred as $expn(E)$ or $expn_\mathcal{A}(E)$. It inherits polarity from $E$ and has also an instantiation function $Inst:(e,\epsilon)\mapsto\epsilon(e)$.
\end{defi}
\begin{prop}The map $red:(e,\epsilon)\mapsto e$ with $Inst$ is an $\mathcal{A}$-strategy.
\end{prop}
$Proof.\ $The map $red$ is a total map of event structure because it preserves causal dependency, consistent sets and it's injective on configurations. Moreover, by definition of $expn(E)$ we have $sort(Inst(e,\epsilon))=sort(\epsilon(e))=sort(var(e))$. Now check receptivity and innocence w.r.t. $\mathcal{A}$:
\begin{itemize}
    \item Consider $x\in\mathcal{C}^\infty(expn(E))$, $e\in E$ and $a\in|\mathcal{A}|$ s.t. $red(x)\stackrel{e}{\lefttail}\ \&\ pol(e)=-\ \&\ var(e)=\alpha\ \&\ sort(a)=sort(\alpha)$, then define $\epsilon:[e]\rightarrow|\mathcal{A}|$ such that $\epsilon(e)=a$ and $\epsilon(e')=\epsilon'(e')$ for $e'\in[e)$ and $(e',\epsilon')\in x$. The map $\epsilon$ is well-defined because for $e'\in[e)$ we have $e'\in red(x)$ by down-closure and $x$ is consistent so there is a unique $(e',\epsilon')\in x$. Then $(e,\epsilon)\in expn(E)$ by construction and so $x\stackrel{(e,\epsilon)}{\lefttail}$ because $red(x)\cup\{e\}$ is a configuration and $e'\le e\Rightarrow\epsilon'=\epsilon\upharpoonright[e']$. Finally $red(e,\epsilon)=e$ and $Inst(e,\epsilon)=a$ give receptivity.
    \item Consider $(e^+,\epsilon^+)\rightarrow(e^-,\epsilon^-)$ in $expn(E)$ s.t. $pol(e^\pm)=\pm$ (that is $pol(e^+)=+$ or $pol(e^-)=-$). Then by definition $e^+\le e^-$ and any element $e$ in between would give $(e,\epsilon^-\upharpoonright[e])$ between $(e^+,\epsilon^+)$ and $(e^-,\epsilon^-)$ so we have $e^+\rightarrow e^-$.
\end{itemize}
\begin{flushright}\qed\end{flushright}
\begin{defi}Local instantiation\\
Consider an $\mathcal{A}$-strategy $\sigma,\ inst:S\rightarrow E,\ |\mathcal{A}|$. For $s\in S$, define by induction the local instantiation function $\epsilon_s:[\sigma(s)]\rightarrow|\mathcal{A}|$ as follows:\\
\begin{center}
    If $[\sigma(s)]=\{\sigma(s)\}$, write $\epsilon_s(\sigma(s))=inst(s)$\\
    If for every $e'\in[\sigma(s))$ there is $s'\in S$ s.t. $\sigma(s')=e'$ and $\epsilon_{s'}$ is defined, write $\epsilon_s(\sigma(s))=inst(s)$ and $\epsilon_s(e')=\epsilon_{s'}(e')=\epsilon_{s'}(\sigma(s'))=inst(s')$
\end{center}
\end{defi}
\begin{lemm}\label{lma1}$\forall s,s'\in S$ s.t. $\sigma(s')\le\sigma(s)$ we have $\epsilon_s\upharpoonright[\sigma(s')]=\epsilon_{s'}$
\end{lemm}
$Proof.\ $By induction on $[\sigma(s)]\subseteq E$:
\begin{itemize}
    \item If $[\sigma(s)]=\{\sigma(s)\}$ then $\sigma(s')\le\sigma(s)\Rightarrow s'=s\Rightarrow\\\epsilon_s\upharpoonright[\sigma(s')]=\epsilon_s\upharpoonright[\sigma(s)]=\epsilon_s=\epsilon_{s'}$.
    \item Suppose $\forall \sigma(s')\in[\sigma(s))$ we have $\sigma(s'')\le\sigma(s')\Rightarrow\epsilon_{s'}\upharpoonright[\sigma(s'')]=\epsilon_{s''}$. Consider $\sigma(s')\in[\sigma(s))$, for every $\sigma(s'')\in[\sigma(s')]$ we have $\epsilon_{s'}(\sigma(s''))=\epsilon_{s''}(\sigma(s''))$ by induction and we have $\epsilon_s(\sigma(s''))=\epsilon_{s''}(\sigma(s''))$ by definition of $\epsilon_s$ with $\sigma(s'')\in[\sigma(s')]\subset[\sigma(s))$. Hence $\sigma(s')\le\sigma(s)\Rightarrow\epsilon_s\upharpoonright[\sigma(s')]=\epsilon_{s'}$.
\end{itemize}
\begin{flushright}\qed\end{flushright}
This is our main theorem, a bijection between the (classical) strategies of $expn(E)$ and the $\mathcal{A}$-strategies of the $\mathcal{A}$-game $(E,\mathcal{A})$.
\begin{prop}$\mathcal{A}$-strategies $\sigma,\ inst$ on $E$ are in one-to-one correspondence with strategies $\sigma':S\rightarrow expn(E)$, i.e.:
\begin{center}
\begin{tikzcd}
S \arrow[dd, "{\sigma,\ inst}" description] &                     & S \arrow[dd, "\sigma'"']                    \\
                                            & \longleftrightarrow &                                             \\
{E,\ |\mathcal{A}|}                           &                     & expn(E) \arrow[ll, "{red,\ Inst}", dotted]
\end{tikzcd}
\end{center}
\end{prop}
$Proof.\ $Consider an $\mathcal{A}$-strategy $\sigma,\ inst:S\rightarrow E,|\mathcal{A}|$, define $\theta(\sigma,inst)=\sigma'$ s.t. $\forall s\in S,\ \sigma'(s)=(\sigma(s),\epsilon_s)$. Let's show that $\sigma'$ is a strategy:\par
Consider $x\in\mathcal{C}^\infty(S)$
\begin{itemize}
    \item Let $Y\subseteq\sigma'x$ s.t. $Y$ is finite and $s_1,s_2\in x$ s.t. $\sigma'(s_i)=(e_i,\epsilon_i)\in Y$. Remark $red(Y)$ is a finite subset of $red(\sigma'x)=\sigma x$ and $\sigma$ is a map of event structures so $red(Y)\in Con_E$. Thus $e_1$ and $e_2$ are consistent, so $\sigma'(s_1)$ and $\sigma'(s_2)$ are consistent iff $e_1=e_2\Rightarrow\forall e'\le e_1,\ \epsilon_1(e')=\epsilon_2(e')$. If $e'\le e_1=e_2$ there is $s'\in S$ s.t. $e'=\sigma(s')\le\sigma(s_1)=\sigma(s_2)$, by \ref{lma1} we have $\epsilon_1\upharpoonright[e']=\epsilon_{s'}=\epsilon_2\upharpoonright[e']$, in particular $\epsilon_1(e')=\epsilon_2(e')$. Hence $Y\in Con_{expn(E)}$.
    \item Let $(e,\epsilon),(e',\epsilon')\in expn(E)$ s.t. $(e',\epsilon')\le(e,\epsilon)$ and $(e,\epsilon)\in\sigma'x$. We have $e\in\sigma x$, $e'\le e$ and $\epsilon'=\epsilon\upharpoonright[e']$ so $e'\in\sigma x$ because $\sigma x$ is down-closed. Thus $\exists s,s'\in x$ s.t. $\sigma'(s)=(e,\epsilon)$ and $\sigma(s')=e'\le\sigma(s)$. Then by Lemma \ref{lma1}, $\epsilon'=\epsilon\upharpoonright[e']=\epsilon_s\upharpoonright[\sigma(s')]=\epsilon_{s'}$ so $\sigma'(s')=(e',\epsilon')$ i.e. $(e',\epsilon')\in\sigma'x$. Hence $\sigma'x$ is down-close.
    \item Let $s_1,s_2\in x$ s.t. $\sigma'(s_1)=\sigma'(s_2)$. Then $\sigma(s_1)=\sigma(s_2)$ and so $s_1=s_2$ because $\sigma$ is locally injective on $x$.
\end{itemize}\par
Thus $\sigma'x\in\mathcal{C}^\infty(expn(E))$ and $\sigma'$ is locally injective on $x$, i.e. $\sigma'$ is a map of event structure. Since $\sigma$ and $s\mapsto\epsilon_s$ are total, $\sigma'$ is too and thus is a pre-strategy.
\begin{itemize}
    \item Consider $x\in\mathcal{C}^\infty(S)$ and $(e,\epsilon)\in expn(E)$ s.t. $\sigma'x\stackrel{(e,\epsilon)}{\lefttail}$ and $pol(e,\epsilon)=-$. By definition we have $\sigma x\stackrel{e}{\lefttail}$, $pol(e)=-$ and $sort(\epsilon(e))=sort(var(e))$. Then by receptivity of $\sigma,inst$ there is a unique $s\in S$ s.t. $x\stackrel{s}{\lefttail}$, $\sigma(s)=e$ and $inst(s)=\epsilon(e)$. Now remark that $\sigma'x\cup\{(e,\epsilon)\}$ and $\sigma'x\cup\{(e,\epsilon_s)\}$ are configurations so for every $e'\not=e,\ e'\le e$ there is $(e',\epsilon'=\epsilon\upharpoonright[e'])$ and $(e',\epsilon_s'=\epsilon_s\upharpoonright[e'])$ consistent in $\sigma'x$, thus $\epsilon'(e')=\epsilon_s'(e')$ i.e. $\epsilon(e')=\epsilon_s(e')$. Hence $\sigma'(s)=(e,\epsilon)$.
    \item Consider $s^+,s^-\in S$ s.t. $s^+\rightarrow s^-$ and $pol(s^\pm)=\pm$ (that is $pol(s^+)=+$ or $pol(s^-)=-$). Then $\sigma(s^+)\rightarrow\sigma(s^-)$ because $\sigma$ is innocent, i.e. $\sigma(s^+)\not=\sigma(s^-)$ and $\sigma(s^+)\le\sigma(s^-)$ with no element in between. Thus $\sigma'(s^+)\not=\sigma'(s^-)$ and $\epsilon_{s^-}\upharpoonright[\sigma(s^+)]=\epsilon_{s^+}$, hence $\sigma'(s^+)\le\sigma'(s^-)$. Any element in between $\sigma'(s^+)$ and $\sigma'(s^-)$ would give an element in between $\sigma(s^+)$ and $\sigma(s^-)$ so there is none, i.e. $\sigma'(s^+)\rightarrow\sigma'(s^-)$.
\end{itemize}\par
So $\sigma'$ is a receptive and innocent pre-strategy, i.e. $\sigma'$ is a strategy.\\\par
Now let's show that $\theta:\sigma,inst\mapsto\sigma'$ is a bijection. First, consider two $\mathcal{A}$-strategies $\sigma_1,\ inst_1$ and $\sigma_2,\ inst_2$ s.t. $\theta(\sigma_1,inst_1)=\theta(\sigma_2,inst_2)$. Then for all $s\in S$ we have $(\sigma_1(s),\epsilon_{1,s})=(\sigma_2(s),\epsilon_{2,s})$ so $inst_1(s)=\epsilon_{1,s}(\sigma_1(s))=\epsilon_{2,s}(\sigma_1(s))=\epsilon_{2,s}(\sigma_2(s))=inst_2(s)$. Thus $(\sigma_1,inst_1)=(\sigma_2,inst_2)$, i.e. $\theta$ is injective. To show that it is surjective we consider a strategy $\sigma':S\rightarrow expn(E)$ and show that $(\sigma,inst)=(red\circ\sigma',Inst\circ\sigma')$ is an $\mathcal{A}$-strategy s.t. $\theta(\sigma,inst)=\sigma'$.\par
Consider $x\in\mathcal{C}^\infty(S)$:
\begin{itemize}
    \item We have $\sigma'x\in\mathcal{C}^\infty(expn(E))$, moreover $(e,\epsilon)$ and $(e',\epsilon')$ consistent implies $e$ and $e'$ consistent and $(e,\epsilon)\le(e',\epsilon')$. Thus $\sigma x\in\mathcal{C}^\infty(E)$.
    \item Let $s_1,s_2\in x$ s.t. $\sigma(s_1)=\sigma(s_2)$, we know that $\sigma'(s_1)$ and $\sigma'(s_2)$ are consistent i.e. $\sigma(s_1)$ and $\sigma(s_2)$ are consistent and $\sigma(s_1)=\sigma(s_2)\Rightarrow\forall e\le\sigma(s_1),\ \epsilon_{s_1}(e)=\epsilon_{s_2}(e)$. Hence $\epsilon_{s_1}=\epsilon_{s_2}$ so $\sigma'(s_1)=\sigma'(s_2)$ and $s_1=s_2$ by local injectivity of $\sigma'$.
\end{itemize}\par
Thus $\sigma$ is a pre-strategy (it is total because $\sigma'$ is). Now we check that $\sigma$ and $inst$ have the right properties:
\begin{itemize}
    \item For $s\in S$, we have $sort(inst(s))=sort(Inst(\sigma'(s)))=sort(\epsilon(e))$ where $(e,\epsilon)=\sigma'(s)\in expn(E)$. Thus $sort(inst(s))=sort(var(e))$ by definition of $expn(E)$ and finally $sort(inst(s))=sort(var(\sigma(s)))$ by definition of $\sigma$.
    \item Consider $x\in\mathcal{C}^\infty(S)$, $e\in E$ and $a\in|\mathcal{A}|$ s.t. $\sigma x\stackrel{e}{\lefttail}\ \&\ pol(e)=-\ \&\ var(e)=\alpha\ \&\ sort(a)=sort(\alpha)$. Define $\epsilon:[e]\rightarrow|\mathcal{A}|$ as $\epsilon(e)=a$ and $\epsilon(e')=\epsilon'(e')$ for all $(e',\epsilon')\in\sigma'x$ s.t. $e'\le e$ from \ref{lma1}. We have $\sigma'x\stackrel{(e,\epsilon)}{\lefttail}$ and $pol(e,\epsilon)=-$ so by receptivity of $\sigma'$: $\exists!s\in S$ s.t. $x\stackrel{s}{\lefttail}$ and $\sigma'(s)=(e,\epsilon)$. Thus $\exists!s\in S$ s.t. $x\stackrel{s}{\lefttail}$, $\sigma(s)=(e)$ and $inst(s)=\epsilon(e)=a$.
    \item Consider $s^+,s^-\in S$ s.t. $s^+\rightarrow s^-$ and $pol(s^\pm)=\pm$. Then $\sigma'(s^+)\rightarrow\sigma'(s^-)$ because $\sigma'$ is innocent, i.e. $\sigma'(s^+)\not=\sigma'(s^-)$ and $\sigma'(s^+)\le\sigma'(s^-)$ with no element in between. Thus $\sigma(s^+)\not=\sigma(s^-)$ (because $e=e'\ \&\ (e,\epsilon)\le(e',\epsilon')\Rightarrow(e,\epsilon)=(e',\epsilon')$) and $\sigma(s^+)\le\sigma(s^-)$ by definition of $\sigma'(s^+)\le\sigma'(s^-)$. Any element in between $\sigma(s^+)$ and $\sigma(s^-)$ would give an element in between $\sigma'(s^+)$ and $\sigma'(s^-)$ so there is none, i.e. $\sigma(s^+)\rightarrow\sigma(s^-)$.
\end{itemize}
\begin{flushright}\qed\end{flushright}
For $\sigma':S\rightarrow expn(E)$ write $reduc(\sigma')$ the unique $\mathcal{A}$-strategy $\sigma,inst$ s.t. $\theta(\sigma,inst)=\sigma'$. Now let's define composition of strategies w.r.t. algebras, first we need to explicit the relation between $expn$ and $||$.
\begin{lemm}Distribution of the expansion w.r.t. two algebras\\
Let $E$ and $F$ be games on the algebras $\mathcal{A}$ and $\mathcal{B}$ respectively. Then,
\begin{align*}
    expn_{\mathcal{A}||\mathcal{B}}(E||F) \cong expn_\mathcal{A}(E)||expn_\mathcal{B}(F)
\end{align*}
\end{lemm}
$Proof.$ The event structure $expn_{\mathcal{A}||\mathcal{B}}(E||F)$ has elements $((1,e),\epsilon:[(1,e)]\rightarrow\{1\}\times|\mathcal{A}|)$ and $((2,f),\phi:[(2,f)]\rightarrow\{2\}\times|\mathcal{B}|)$ with $e\in E,\ f\in F$ ($\epsilon$ and $\phi$ are restricted to one algebra since the there is no causal dependency between $E$ and $F$ in $E||F$). So we can easily project these elements to $(1,(e,\pi_2\epsilon\pi_2:[e]\rightarrow|\mathcal{A}|)$ and $(2,(f,\pi_2\phi\pi_2:[f]\rightarrow|\mathcal{B}|)$ for $e\in E,\ f\in F$ and vice versa. Hence the isomorphism.\begin{flushright}\qed\end{flushright}
\begin{defi}Composition of winning strategies\\
Consider a strategy $\sigma,inst_\sigma:S\rightarrow E^\bot||F$ and a strategy $\tau,inst_\tau:S\rightarrow F^\bot||G$. Define the (synchronized) composition of $\sigma$ and $\tau$ as the strategy $\tau\circledast\sigma,inst_\circledast:T\odot S\rightarrow E^\bot||G$ s.t. $\tau\circledast\sigma,inst_\circledast=reduc(\tau'\odot\sigma')$.\\
\end{defi}
\begin{conj}Stability by composition\\
If $\sigma$ and $\tau$ are winning strategies then $\tau\circledast\sigma$ is winning too. That is if $\sigma$ is a strategy from $E$ to $F$ and $\tau$ from $F$ to $G$ then $\tau\circledast\sigma$ is a strategy from $E$ to $G$.
\end{conj}
\begin{tikzcd}
S_{E^\bot} \arrow[d, "\sigma_{E^\bot}"', dashed] & T\odot S \arrow[d, "\tau\circledast\sigma"'] \arrow[l, "p_{E^\bot}\pi_1"', harpoon', dashed] \arrow[r, "p_G\pi_2", harpoon, dashed] & T_G \arrow[d, "\tau_G", dashed] \\
E^\bot                         & E^\bot||G \arrow[r, harpoon] \arrow[l, harpoon']                                            & G                       
\end{tikzcd}
\section{Further results on games}
\subsection{An approach with neutral events}
We extend event structures' polarity with a third value representing moves that are neither from the Player nor the Opponent, i.e. these events can be thought as neutral information coming from the system the two agents are using. An event structure with neutral events is an event structure $S$ with a polarity function $pol : S\rightarrow\{+,-,0\}$. Similarly, maps are maps of event structures which preserve polarity when defined. But in a game with neutral events, Player's and Opponent's moves are independent from neutral ones.
\begin{defi}Game with neutral events on an algebra\\
W.r.t. a relational many-sorted algebra $\mathcal{A}$ with signature $\Sigma$, define an $\mathcal{A}$-game with neutral events to comprise
\begin{align*}
    (\mathcal{A},\ N||E,\ pol,\ var)
\end{align*}
where $(\mathcal{A},E,pol_E,var_E)$ is a regular $\mathcal{A}$-game and $N$ is an event structure with
\begin{align*}
    pol :\ &n\in N\mapsto0\\
    &e\in E\mapsto pol_E(e)\in\{+,-\}
\end{align*}
and $var$ respects (\textbf{no-overlap}) for $N||E$. Finally the $\mathcal{A}$-game with neutral event inherits the winning condition from the $\mathcal{A}$-game $(\mathcal{A},E,pol_E,var_E)$.
\end{defi}
Remark that since $N$ and $E$ are in parallel composition we have $n\ co\ e$ for all $n\in N$ and $e\in E$ because $\{n\}\in Con_N\wedge\{e\}\in Con_E\Rightarrow\{n,e\}\in Con_{N||E}$. And so by (\textbf{no-overlap}) we have $var(n)\not=var(e)$, i.e. $N$ and $E$ use different sets of variables in $Var_\mathcal{A}$. Now let's define strategies for this game.
\begin{defi}Strategy in an $\mathcal{A}$-game with neutral moves\\
An $\mathcal{A}$-strategy in the game $N||E$ comprises a total map of event structure with extended polarity
\begin{align*}
    &S\xrightarrow{\text{  $\sigma$ }}N||E\text{ with}\\
    inst:S\rightarrow|\mathcal{A}|\text{ s.t. }&\forall s\in S,\ sort(inst(s))=sort(var(\sigma(s)))
\end{align*}
which is \textbf{receptive}:
\begin{align*}
    &\forall x\in\mathcal{C}^\infty(S),\ \forall e\in E,\ \sigma x\stackrel{e}{\lefttail}\ \&\ pol(e)=-\ \&\ var(e)=\alpha,\\
    &\forall a\in|\mathcal{A}|,\ sort(a)=sort(\alpha),\\
    &\exists!s\in S,\ x\stackrel{s}{\lefttail}\ \&\ \sigma(s)=e\ \&\ inst(s)=a
\end{align*}
and \textbf{innocent/courteous}:
\begin{align*}
    &s\rightarrow s'\ \&\ pol(s')=-\Rightarrow\sigma(s)\rightarrow\sigma(s')\\
    &s\rightarrow s'\ \&\ pol(s)=+\Rightarrow\sigma(s)\rightarrow\sigma(s')
\end{align*}
Remark that for receptivity we need $pol(e)=-$ which gives $e\not\in N$, also $s$ and $s'$ can be neutral for $-$-innocence and $+$-innocence respectively. These conditions imply that $\sigma_\pm :S_\pm\rightarrow E$ is a strategy on a regular $\mathcal{A}$-game.
\end{defi}
Now the same way we proceeded before, let's define an $\mathcal{A}$-expansion to get rid of variables:
an $\mathcal{A}$-expansion with neutral moves $expn(N||E,\mathcal{A},var)$ is the parallel composition $expn(N,\mathcal{A},var_N)||expn(E,\mathcal{A},var_E)$.
\begin{prop}$\mathcal{A}$-strategies with neutral moves $\sigma,\ inst$ on $N||E$ are in one-to-one correspondence with strategies $\sigma':S\rightarrow expn(N||E)$, i.e.:
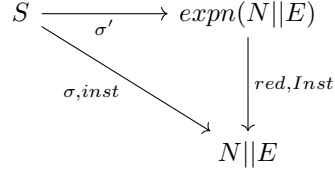
\begin{center}
\begin{tikzcd}
S \arrow[rrdd, "{\sigma, inst}"'] \arrow[rr, "\sigma'"'] &  & expn(N||E) \arrow[dd, "{red, Inst}"] \\
                                                         &  &                                     \\
                                                         &  & N||E                               
\end{tikzcd}
\end{center}
\end{prop}
$Proof.\ $Consider an $\mathcal{A}$-strategy $\sigma,\ inst:S\rightarrow N||E$, define $\theta(\sigma,inst)=\sigma'$ such that $\forall s\in S,\ \sigma'(s)=(\sigma(s),\epsilon_s)$. Let's show that $\sigma'$ is a strategy:\par
Consider $x\in\mathcal{C}^\infty(S)$
\begin{itemize}
    \item Let $Y\subseteq\sigma'x$ s.t. $Y$ is finite and $s_1,s_2\in x$ s.t. $\sigma'(s_i)=(e_i,\epsilon_i)\in Y$. Remark $red(Y)$ is a finite subset of $red(\sigma'x)=\sigma x$ and $\sigma$ is a map of event structures so $red(Y)\in Con_{N||E}$. Thus $e_1$ and $e_2$ are consistent, so $\sigma'(s_1)$ and $\sigma'(s_2)$ are consistent iff $e_1=e_2\Rightarrow\forall e'\le e_1,\ \epsilon_1(e')=\epsilon_2(e')$. If $e'\le e_1=e_2$ there is $s'\in S$ s.t. $e'=\sigma(s')\le\sigma(s_1)=\sigma(s_2)$, by \ref{lma1} we have $\epsilon_1\upharpoonright[e']=\epsilon_{s'}=\epsilon_2\upharpoonright[e']$, in particular $\epsilon_1(e')=\epsilon_2(e')$. Hence $Y\in Con_{expn(N||E)}$.
    \item Let $(e,\epsilon),(e',\epsilon')\in expn(N||E)$ s.t. $(e',\epsilon')\le(e,\epsilon)$ and $(e,\epsilon)\in\sigma'x$. We have $e\in\sigma x$, $e'\le e$ and $\epsilon'=\epsilon\upharpoonright[e']$ so $e'\in\sigma x$ because $\sigma x$ is down-closed. Thus $\exists s,s'\in x$ s.t. $\sigma'(s)=(e,\epsilon)$ and $\sigma(s')=e'\le\sigma(s)$. Then by \ref{lma1}, $\epsilon'=\epsilon\upharpoonright[e']=\epsilon_s\upharpoonright[\sigma(s')]=\epsilon_{s'}$ so $\sigma'(s')=(e',\epsilon')$ i.e. $(e',\epsilon')\in\sigma'x$. Hence $\sigma'x$ is down-close.
    \item Let $s_1,s_2\in x$ s.t. $\sigma'(s_1)=\sigma'(s_2)$. Then $\sigma(s_1)=\sigma(s_2)$ and so $s_1=s_2$ because $\sigma$ is locally injective on $x$.
\end{itemize}\par
Thus $\sigma'x\in\mathcal{C}^\infty(expn(N||E))$ and $\sigma'$ is locally injective on $x$, i.e. $\sigma'$ is a map of event structure. Since $\sigma$ and $s\mapsto\epsilon_s$ are total, $\sigma'$ is too and thus is a pre-strategy.
\begin{itemize}
    \item Consider $x\in\mathcal{C}^\infty(S)$ and $(e,\epsilon)\in expn(E)$ s.t. $\sigma'x\stackrel{(e,\epsilon)}{\lefttail}$ and $pol(e,\epsilon)=-$ (we don't consider $(e,\epsilon)\in expr(N)$ because of polarity). By definition we have $\sigma x\stackrel{e}{\lefttail}$, $pol(e)=-$ and $sort(\epsilon(e))=sort(var(e))$. Then by receptivity of $\sigma,inst$ there is a unique $s\in S$ s.t. $x\stackrel{s}{\lefttail}$, $\sigma(s)=e$ and $inst(s)=\epsilon(e)$. Now remark that $\sigma'x\cup\{(e,\epsilon)\}$ and $\sigma'x\cup\{(e,\epsilon_s)\}$ are configurations so for every $e'\not=e,\ e'\le e$ there is $(e',\epsilon'=\epsilon\upharpoonright[e'])$ and $(e',\epsilon_s'=\epsilon_s\upharpoonright[e'])$ consistent in $\sigma'x$, thus $\epsilon'(e')=\epsilon_s'(e')$ i.e. $\epsilon(e')=\epsilon_s(e')$. Hence $\sigma'(s)=(e,\epsilon)$.
    \item Innocence proof is identical to the case without neutral events.
\end{itemize}\par
So $\sigma'$ is a receptive and innocent pre-strategy, i.e. $\sigma'$ is a strategy.\\\par
Now let's show that $\theta:\sigma,inst\mapsto\sigma'$ is a bijection. First, consider two $\mathcal{A}$-strategies with neutral moves $\sigma_1,\ inst_1$ and $\sigma_2,\ inst_2$ s.t. $\theta(\sigma_1,inst_1)=\theta(\sigma_2,inst_2)$. Then for all $s\in S$ we have $(\sigma_1(s),\epsilon_{1,s})=(\sigma_2(s),\epsilon_{2,s})$ so $inst_1(s)=\epsilon_{1,s}(\sigma_1(s))=\epsilon_{2,s}(\sigma_1(s))=\epsilon_{2,s}(\sigma_2(s))=inst_2(s)$. Thus $(\sigma_1,inst_1)=(\sigma_2,inst_2)$, i.e. $\theta$ is injective. To show that it is surjective we consider a strategy $\sigma':S\rightarrow expn(N||E)$ and show that $(\sigma,inst)=(red\circ\sigma',Inst\circ\sigma')$ is an $\mathcal{A}$-strategy s.t. $\theta(\sigma,inst)=\sigma'$.\par
Consider $x\in\mathcal{C}^\infty(S)$:
\begin{itemize}
    \item We have $\sigma'x\in\mathcal{C}^\infty(expn(N||E))$, moreover $(e,\epsilon)$ and $(e',\epsilon')$ consistent implies $e$ and $e'$ consistent and $(e,\epsilon)\le(e',\epsilon')$. Thus $\sigma x\in\mathcal{C}^\infty(E)$.
    \item Let $s_1,s_2\in x$ s.t. $\sigma(s_1)=\sigma(s_2)$, we know that $\sigma'(s_1)$ and $\sigma'(s_2)$ are consistent i.e. $\sigma(s_1)$ and $\sigma(s_2)$ are consistent and $\sigma(s_1)=\sigma(s_2)\Rightarrow\forall e\le\sigma(s_1),\ \epsilon_{s_1}(e)=\epsilon_{s_2}(e)$. Hence $\epsilon_{s_1}=\epsilon_{s_2}$ so $\sigma'(s_1)=\sigma'(s_2)$ and $s_1=s_2$ by local injectivity of $\sigma'$.
\end{itemize}\par
Thus $\sigma$ is a pre-strategy (it is total because $\sigma'$ is). Now we check that $\sigma$ and $inst$ have the right properties:
\begin{itemize}
    \item For $s\in S$, we have $sort(inst(s))=sort(Inst(\sigma'(s)))=sort(\epsilon(e))$ where $(e,\epsilon)=\sigma'(s)\in expn(N||E)$. Thus $sort(inst(s))=sort(var(e))$ by definition of $expn(N||E)$ and finally $sort(inst(s))=sort(var(\sigma(s)))$ by definition of $\sigma$.
    \item Consider $x\in\mathcal{C}^\infty(S)$, $e\in E$ and $a\in|\mathcal{A}|$ s.t. $\sigma x\stackrel{e}{\lefttail}\ \&\ pol(e)=-\ \&\ var(e)=\alpha\ \&\ sort(a)=sort(\alpha)$ (we don't consider $e\in N$ because of polarity). Define $\epsilon:[e]\rightarrow|\mathcal{A}|$ as $\epsilon(e)=a$ and $\epsilon(e')=\epsilon'(e')$ for all $(e',\epsilon')\in\sigma'x$ s.t. $e'\le e$ \ref{lma1}. We have $\sigma'x\stackrel{(e,\epsilon)}{\lefttail}$ and $pol(e,\epsilon)=-$ so by receptivity of $\sigma'$: $\exists!s\in S$ s.t. $x\stackrel{s}{\lefttail}$ and $\sigma'(s)=(e,\epsilon)$. Thus $\exists!s\in S$ s.t. $x\stackrel{s}{\lefttail}$, $\sigma(s)=(e)$ and $inst(s)=\epsilon(e)=a$.
    \item Innocence proof is identical to the case without neutral events.
\end{itemize}
\begin{flushright}\qed\end{flushright}
\subsection{Concurrent copycat strategies}
Identities on games are given by copycat strategies: strategies for Player based on copying the latest moves made by Opponent.\\
Let $A$ be an event structure with polarity. The copycat strategy from $A$ to $A$ is an instance of a pre-strategy, so a total map $\gamma_A\ \colon\ CC_A \rightarrow A^\bot||A$. It describes a concurrent, or distributed, strategy based on the idea that Player moves always copy previous corresponding moves of Opponent with opposite polarity. The following proposition is from the Ecsym notes \cite{ecsym}:
\begin{prop}
Let $A$ be an event structure with polarity. There is an event structure with polarity $CC_A$ having the same events and polarity as $A^\bot||A$ but with causal dependency $\leq_{CC_A}$ given as the transitive closure of the relation
\begin{align*}
    \leq_{A^\bot||A}\cup\ \{(\overline{c},c)\ |\ c\in A^\bot||A\ \&\ pol_{A^\bot||A}(c)=+\}
\end{align*}
and finite subsets of $CC_A$ consistent if their down-closure w.r.t. $\leq_{CC_A}$ are consistent in $A^\bot||A$. Moreover,\\
$(i)\ c\rightarrow c'\ in\ CC_A\ iff$
\begin{align*}
    c\rightarrow c'\ in\ A^\bot||A\ or\ pol_{A^\bot||A}(c')=+\ \&\ \overline{c}=c'
\end{align*}
$(ii)\ x\in\mathcal{C}(CC_A)\ iff$
\begin{align*}
    x\in\mathcal{C}(A^\bot||A)\ \&\ \forall c\in x,\ pol_{A^\bot||A}(c)=+\Rightarrow \overline{c}\in x
\end{align*}
\end{prop}
Based on this proposition define the copycat pre-strategy from $A$ to $A$ to be the pre-strategy $\gamma_A\ \colon\ CC_A\rightarrow A^\bot||A$ where $CC_A$ comprises the event structure with polarity $A^\bot||A$ together with extra causal dependencies $\overline{c}\leq_{CC_A}c$ for all events $c$ with $pol_{A^\bot||A}(c)=+$, and $\gamma_A$ is the identity on the set of events common to both $CC_A$ and $A^\bot||A$.\par
We can extend this definition for $\mathcal{A}$-games and try to obtain a bicategory of concurrent games with algebras.
\begin{defi}Copycat strategy of a game on an algebra\\
For an $\mathcal{A}$-game $(E,\mathcal{A},var)$ let the copycat strategy w.r.t. to algebra $\mathcal{A}$ be the composition:
\begin{center}
\begin{tikzcd}
CC_{expn(E)} \arrow[rr, "\gamma_{expn(E)}"] &  & expn(E)^\bot||expn(E) \arrow[rr, "ep^\bot||ep"] &  & E^\bot||E
\end{tikzcd}
\end{center}
\end{defi}
\subsection{$\mathcal{A}$-algebras from $\mathcal{A}$-strategies and games}
\begin{defi}
Let $\sigma:S\rightarrow E$ be an $\mathcal{A}$-strategy. Then $S$ inherits the structure of an $\mathcal{A}$-algebra. For $R$ in the signature define an $\mathcal{A}$-algebra on events in $S$ by $R(s_1,...,s_k)$ if and only if:
\begin{align*}
     \{s_1,...,s_k\} &\text{ is consistent in } S\\
     \forall i,\ 1\le i\le k,\ \exists\alpha\in V&ar_\mathcal{A}\text{ s.t. }\sigma(s_i)=latest(\alpha,\sigma x)\\
     R^\mathcal{A}(inst&(s_i),...,inst(s_k))
\end{align*}
where $x:=[s_1,...,s_k]$
\end{defi}
In particular an $\mathcal{A}$-game $E$ determines an $\mathcal{A}$-strategy $ep:expn(E)\rightarrow E$ and so an $\mathcal{A}$-algebra on the events of $expn(E)$.
\begin{conj}
$R(s_1,...,s_k)$ iff $\exists x\in\mathcal{C}(S),\ v\in\mathbb{N}^k$ s.t.
\begin{align*}
     s_1,...,s_k\in\ &x\\
     x\models\mathbb{E}(\alpha_{v(1)})\wedge...\wedge\mathbb{E}(\alpha_{v(k)})&\wedge R(\alpha_{v(1)},...,\alpha_{v(k)})\\
     \forall i,\ 1\le i\le k,\ \sigma(s_i)=l&atest(\alpha_{v(i)},\sigma x)
\end{align*}
\end{conj}
\section{Example of games}
\subsection{$A$-homomorphism game}
Let $\mathcal{A}$ and $\mathcal{B}$ be single-sorted algebras over a common signature $\Sigma$. Assume a countable bijection $(\alpha_n,\beta_n)_{n\in\mathbb{N}}$ between the variables $\alpha\in Var_\mathcal{A}$ and $\beta\in Var_\mathcal{B}$. Build the homomorphism game over algebra $\mathcal{A}||\mathcal{B}$ as follows. Its events comprise non-empty sequences $\alpha_{i_1},\beta_{i_1},\alpha_{i_2},\beta_{i_2},...,\alpha_{i_n},\beta_{i_n}$ and $\alpha_{i_1},\beta_{i_1},\alpha_{i_2},\beta_{i_2},...,\alpha_{i_n}$ with causal dependency given by initial sub-sequences, conflict by branching, polarity + for even length and - for odd length, and the $var$ function given by taking the last variable of the sequence.\\
The winning condition makes use of a countably infinite conjunction:
\begin{align*}
    \bigwedge_{i,j\in\mathbb{N}}\alpha_i=\alpha_j\rightarrow\beta_i=\beta_j\ \ \wedge\bigwedge_{R\in\Sigma_k,v\in\mathbb{N}^k}R^\mathcal{A}(\alpha_{v_1},...,\alpha_{v_k})\rightarrow R^\mathcal{B}(\beta{v_1},...,\beta{v_k})
\end{align*}
\begin{conj}
If $\mathcal{A}$ is countable, deterministic winning strategies correspond to homomorphisms from $\mathcal{A}$ to $\mathcal{B}$
\end{conj}
We obtain a $k$-pebbled version of the game by restricting to variables in the finite bijection $(\alpha_1,\beta_1),...,(\alpha_k,\beta_k)$.
\subsection{Ehrenfencht-Fraissé games}
Similar to above but where events are sequences $\gamma_{i_1},\gamma'_{i_1},\gamma_{i_2},\gamma'_{i_2},...,\gamma_{i_n},\gamma'_{i_n}$ and $\gamma_{i_1},\gamma'_{i_1},\gamma_{i_2},\gamma'_{i_2},...,\gamma_{i_n}$ where $\forall k,\ (\gamma_{i_k},\gamma'_{i_k})=(\alpha_{i_k},\beta_{i_k})$ or $(\beta_{i_k},\alpha_{i_k})$. The winning condition is:
\begin{align*}
    \bigwedge_{i,j\in\mathbb{N}}\alpha_i=\alpha_j\iff\beta_i=\beta_j\ \ \wedge\bigwedge_{R\in\Sigma_k,v\in\mathbb{N}^k}R^\mathcal{A}(\alpha_{v_1},...,\alpha_{v_k})\iff R^\mathcal{B}(\beta{v_1},...,\beta{v_k})
\end{align*}
\begin{conj}
If $\mathcal{A}$ and $\mathcal{B}$ are countable, deterministic winning strategies correspond to isomorphisms from $\mathcal{A}$ to $\mathcal{B}$
\end{conj}
We look now to see if we can obtain such games as games $\mathcal{G}(\mathcal{A})^\bot\upand\mathcal{G}(\mathcal{B})$ for a construction $\mathcal{G}$ of games from algebras. To stand a chance of doing so we seem to need a new means of restricting strategies. We try via games of imperfect information.
\section{An extension to imperfect information}
\subsection{Definition of access levels}
Let $(\Lambda,\lesssim)$ be a preorder of access levels.\par
A $\Lambda$-game is a game $(A,pol)$ with a function $l:A\rightarrow\Lambda$ s.t. $a'\leq_Aa\Rightarrow l(a')\lesssim l(a)$ for all $a,a'\in A$. A $\Lambda$-strategy is a strategy $\sigma:S\rightarrow A$ s.t. $s'\leq_Ss\Rightarrow l\sigma(s')\lesssim l\sigma(s)$ for all $s,s'\in S$.
\begin{defi}Operations on $\Lambda$-games\\
We assume a fixed $\Lambda$ throughout:
\begin{align*}
    (A,l)^\bot&=(A^bot,l)\\
    (A,l_A)\upand(B,l_B)&=(A||B,l_A||l_B)
\end{align*}
A $\Lambda$-strategy $(A,l_A)\nvrightarrow(B,l_B)$ is a strategy in $(A,l_A)^\bot\upand(B,l_B)$. For a $\Lambda$-game $(A,l)$, the $\Lambda$-copycat is the copycat strategy $cc_A:CC_A\rightarrow A^\bot||A$ with $l_{CC_A}(c)=l_{A^\bot||A}(cc_A(c))$ for all $c\in CC_A$.
\end{defi}
From \cite{ecsym} we have the following result:
\begin{theo}
$\Lambda$-strategies compose with $\Lambda$-copycat as identity
\end{theo}
We can combine accessibility $\Lambda$ with the structure of $\mathcal{A}$-games and strategies. For a many-sorted algebra $\mathcal{A}$, an $\mathcal{A},\Lambda$-game is an $\mathcal{A}$-game $(A,var)$ together with $l:A\rightarrow\Lambda$ s.t. $a'\leq a\Rightarrow l(a')\lesssim l(a)$ for all $a,a'\in|\mathcal{A}|$. An $\mathcal{A},\Lambda$-strategy is an $\mathcal{A}$-strategy $\sigma:S\rightarrow A,inst$ s.t. $S\xrightarrow{\text{  $\sigma$ }}A\xrightarrow{\text{  $ep$ }}expn(A)$ is a $\Lambda$-strategy ; we understand $expn(A)$ as a $\Lambda$-game with labelling $l_{expn(A)}(e)=l(ep(e))$ for $e\in expn(A)$.\par
In the following we shall assume a fixed preorder of access levels $\Lambda$, that all games are $\Lambda$-games and so we will put aside explicit mentions of $\Lambda$. Though it will, of course, affect the nature of strategies, possibly forbidding causal connections that might otherwise be there.
\begin{rema}
We could perhaps simplify the definition of $\mathcal{A},\Lambda$-games by assuming that the variables carry access levels in $\Lambda$, thus inducing a level function on the game.
\end{rema}
\begin{prop}
Let $\sigma:S\rightarrow E$ be an $\mathcal{A},\Lambda$-strategy. Suppose $e,e'\in E$ are $\Lambda$-independent, i.e. $l(e)$ and $l(e')$ are incomparable, with $pol(e)=-$, $pol(e')=+$, $\alpha=var(e)$ and $\alpha'=var(e')$. Finally let $x\in\mathcal{C}^\infty(S)$ s.t. $\sigma x\stackrel{e}{\lefttail}$ and $\sigma x\stackrel{e'}{\lefttail}$, then:
\begin{align}
    &x\stackrel{s}{\lefttail}\wedge\  x\stackrel{s'}{\lefttail}\wedge\ \sigma(s)=e\wedge\sigma(s')=e'\Rightarrow x\cup\{s,s'\}\in\mathcal{C}^\infty(S)\\
    &\left\{ (\alpha,\alpha')[x']\ \middle\vert \begin{array}{l}
    \exists s,s'\in S,\sigma(s)=e\wedge\sigma(s')=e'\wedge\\
    x\stackrel{s}{\lefttail}\wedge\  x\stackrel{s'}{\lefttail}\wedge\ x'=x\cup\{s,s'\}
    \end{array}\right\}=\nonumber\\
    &\quad\big\{\alpha[x']\mid\exists s\in S,\sigma(s)=e\wedge x\stackrel{s}{\lefttail}\wedge\ x'=x\cup\{s\}\big\}\times\\
    &\quad\big\{\alpha'[x']\mid\exists s'\in S,\sigma(s')=e'\wedge x\stackrel{s'}{\lefttail}\wedge\ x'=x\cup\{s'\}\big\}\nonumber
\end{align}
\end{prop}\par\ \par
$Proof.\ $(1) follows by receptivity and (2) as a consequence of (1)\begin{flushright}\qed\end{flushright}
\subsection{Techniques for reasoning about $\mathcal{A},\Lambda$-strategies}\setcounter{equation}{0}
Generally, for a strategy $\sigma:S:\rightarrow E$ we have
\begin{align}
    \begin{tikzcd}
x\in\mathcal{C}^\infty(S) \arrow[d, maps to] \\
\sigma x\subseteq^-y                        
\end{tikzcd}\Rightarrow\begin{tikzcd}
\exists!x'\geq x \arrow[d, "\sigma", maps to] \\
y=\sigma x'                                  
\end{tikzcd}\\
\begin{tikzcd}
x\in\mathcal{C}^\infty(S) \arrow[d, maps to] \\
y\subseteq^+\sigma x                       
\end{tikzcd}\Rightarrow\begin{tikzcd}
\exists!x'\leq x \arrow[d, "\sigma", maps to] \\
y=\sigma x'                                  
\end{tikzcd}
\end{align}
With imperfect information we can do more, e.g. if $x\in\mathcal{C}^\infty(S),e\in E$ s.t. $y\stackrel{e}{\lefttail}\sigma x$, $pol(e)=+$ and $\forall e'\in\sigma x,\ pol(e')=+\Rightarrow l(e)\not\lesssim l(e')$, then $\exists!s',\ x'\in\mathcal{C}^\infty(S)$ s.t. $x'\stackrel{s'}{\lefttail}x$ and $\sigma x'=y$.\\
Moreover then we could expend $x'$ to $x''\in\mathcal{C}^\infty(S)$ with $x'\stackrel{s''}{\lefttail}x''$ associated with another choice for $var(e)$. Under appropriate assumptions on $\Lambda$ we can swap $\mathcal{A}$-values assigned to negative moves by the strategy. Note that in doing so we may lose $+$-maximality.
\section{Conclusion}
What has been achieved is showing a bijection between strategies on a game with algebra and strategies on a regular - albeit more complex - game. We also succeeded in showing the results holds with neutral events.\par
However it is still unclear if a well-formed bicategory of concurrent games with algebras can be defined. Our attempts to compose these strategies while managing the added structure didn't pan out.\par
Concerning the other classic extensions of concurrent games the first results we presented show promise of a more general usage of games with algebra.
\newpage
\bibliographystyle{plain}
\bibliography{references}
\end{document}